\documentclass[journal]{IEEEtran}

\usepackage{verbatim}
\usepackage{rotating}
\usepackage{mathrsfs}
\usepackage{stmaryrd}
\usepackage{tikz}

\usetikzlibrary{arrows}
\usetikzlibrary{calc}
\usepackage{pgfplots}
\usepackage{amsmath,amssymb}
\usepackage{booktabs}
\usepackage{makecell}
\usepackage{float}
\usetikzlibrary{positioning}
\usepackage[ruled,linesnumbered]{algorithm2e}

\usetikzlibrary{external}
\tikzsetfigurename{tikz_compiled} 

{}
{}
{}

\begin{document}


\title{Complexity-Adjustable SC Decoding of Polar Codes for Energy Consumption Reduction}

\author{Haotian Zheng, Bin Chen, Luis F. Abanto-Leon, Zizheng Cao, Ton Koonen

\thanks{H. Zheng, Z. Cao, and A.~M.~J.~Koonen are with the Department of Electrical Engineering, Eindhoven University of Technology, The Netherlands (\mbox{e-mails:} z.cao@tue.nl).}

\thanks{B. Chen was with the Department of Electrical Engineering, Eindhoven University of Technology, The Netherlands and is now with the School of Computer and Information, Hefei University of Technology (HFUT), Hefei, China.}

\thanks{Luis F. Abanto-Leon is with the Department of Computer Science, Technische Universität Darmstadt, Darmstadt, Germany.}

}


\maketitle

\begin{abstract}
This paper proposes an enhanced list-aided successive cancellation stack (ELSCS) decoding algorithm with adjustable decoding complexity. In addition, a logarithmic likelihood ratio (LLR)-threshold based path extension scheme is designed to further reduce the memory consumption of stack decoding. Numerical simulation results show that without affecting the error correction performance, the proposed ELSCS decoding algorithm provides a flexible tradeoff between time complexity and computational complexity, while reducing storage space up to $70\%$. Based on the fact that most mobile devices operate in environments with stringent energy budget to support diverse applications, the proposed scheme is a promising candidate for meeting requirements of different applications while maintaining a low computational complexity and computing resource utilization.
\end{abstract}


\section{Introduction}\label{sec:intro}

With the ever-increasing number of mobile communication devices and amount of communication data, power consumption becomes a tremendous challenge for mobile communication devices, especially for those operating under very stringent energy budget. To obtain high energy efficiency, the high energy-consuming error-correcting code (ECC) decoder is a pivotal component in the communications chain that needs to be suitably designed. However, different applications have diverse set of communication requirements, an one-size-fits-all energy-efficiency decoder is unfeasible. Therefore, a flexible decoder that can adjust the tradeoff between the conflicting demands for energy, throughput and error rate performance is essential for achieving low energy expenditure in different applications.\par

Considering its explicit decoding construction and low decoding complexity, polar codes are strong candidates to fulfill such requirements. These codes were proposed by Arikan and are the first provable capacity-achieving code for any binary-input discrete memoryless channels \cite{arikan2009,arikan2009rate}. Polar codes have been shown to outperform LDPC code at moderate and short code length \cite{tal2015list,koike2018irregular}. In the newest 5G coding standard, polar codes have been adopted as channel coding for control channels in the enhanced mobile broadband (eMBB) communications services \cite{3GPP}.\par
The most widely used high-performance decoding algorithm for polar codes is CRC-aided successive cancellation list (CA-SCL) decoding \cite{tal2015list,balatsoukas2015llr}. Nevertheless, it still requires a large list size to achieve a good error rate performance, which inevitably leads to excessive computational complexity. Another drawback is that the complexity of CA-SCL remains constant in different signal-to-noise ratio (SNR) regimes when in fact fewer computations are required for correct decoding in the high SNR regime, thus causing unnecessary energy expenditure.\par
   Previous works have studied the simplified calculation of successive cancellation (SC)/successive cancellation list (SCL) decoding \cite{alamdar2011simplified,sarkis2013increasing,sarkis2014fast,sarkis2014increasing,sarkis2016fast}. They found removable and replaceable redundant calculations when encountering frozen bit and particular information patterns, while keeping the error rate performance unaltered.  With a similar principle, a new class of relaxed polar codes was introduced by deactivating polarization units when the bit-channels are sufficiently good or bad \cite{el2017relaxed}. Irregular polar codes generalized this idea to further consider all possible selections of the inactivated polarization units and obtained a significant decoding complexity reduction \cite{koike2018irregular}.\par
   In addition to simplification of the decoding process, path pruning is also an efficient way to reduce the computational complexity of SCL decoding. K. Chen et al. proposed to set up a flexible path metric threshold to prune decoding paths with small path metric \cite{chen2016reduce,chen2013improved}. In \cite{chen2016low}, the concept of relative path metric (RPM) was introduced and the correct candidate was found to have a low probability of having a large RPM value. Paths which do not satisfy this property are pruned and the computational complexity can thereby be effectively decreased.\par
   Adaptive SCL algorithm was introduced in \cite{li2012adaptive}. It sets the initial value of the list size to 1. If the decoding attempt succeeds, the decoding process stops and the decoding result is output. Otherwise, the list size is doubled and the SCL decoding algorithm is restarted. This procedure is repeated until the list size reaches a predefined maximum threshold. If the decoding attempt still fails, a decoding failure is declared. This tentative early-stopped strategy enables adaptive SCL algorithm to have SNR-adaptive capability. Furthermore, this kind of strategy is also applicable to other decoding algorithms. As long as the algorithm itself has a lower computational complexity than SCL, the resultant decoding complexity of the new hybrid using this strategy will also be lower.\par
In fact, when decoding latency is not so stringent and throughput demand is not very high, another successive cancellation stack (SCS) decoding algorithm \cite{niu2012stack} is a more energy-efficient choice than SCL. By means of trading storage complexity for computational complexity reduction, SCS has a much lower and SNR-adaptive computational complexity. It was employed inside the kernel processor to reduce the decoding complexity of polar codes with Reed-Solomon kernel substantially \cite{Trifonov2018}. In addition, the efficient
software implementation of the SCS decoding algorithm has been investigated \cite{Aurora2018}. Moreover, the computational complexity reduction schemes mentioned above are equally applicable to SCS, can further decrease the required amount of computations. However, lower energy consumption leads to the disadvantage of smaller throughput. Because of employing breadth-first instead of depth-first search, the decoding delay of SCS will be larger than that of SCL with the same error rate performance, especially in the low SNR regime.\par 

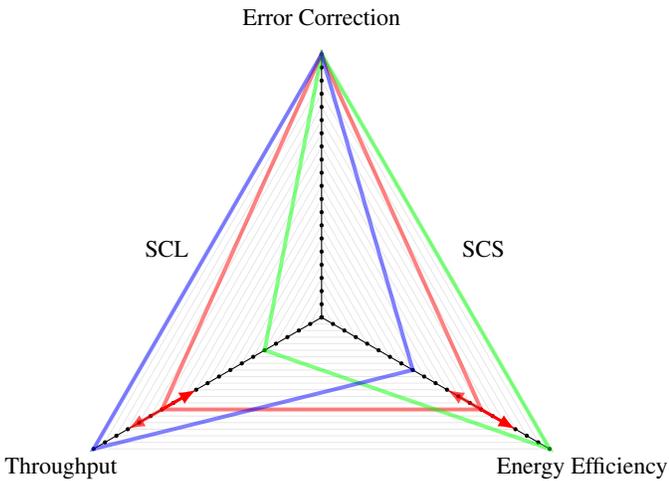
\begin{figure}[!htbp]
\setlength{\abovecaptionskip}{1cm}
\centering
\includegraphics{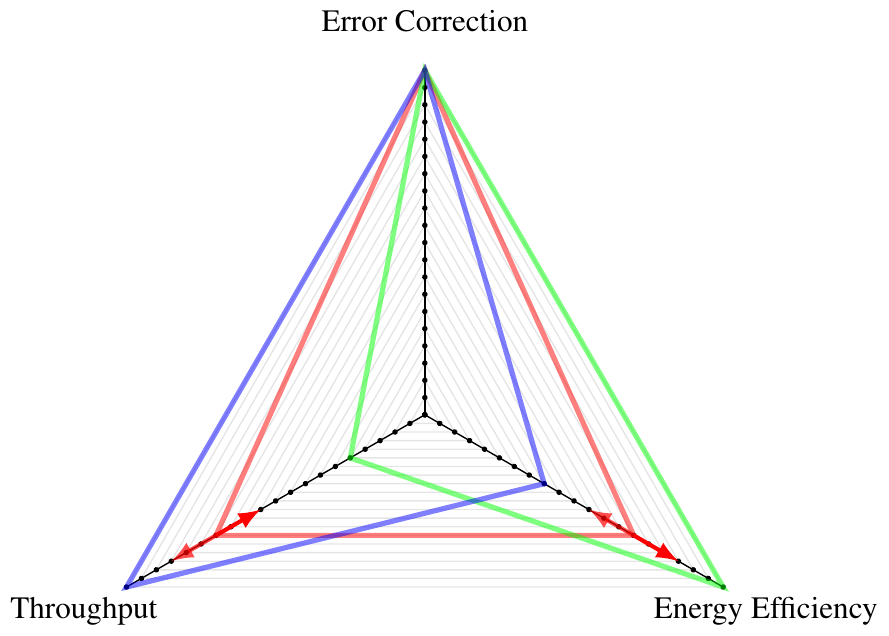}
  \begin{picture}(0,0)
    \put(-70,99){\small SCL}
    \put(50,99){\small SCS}
  \end{picture}
\caption{Comparison of performance metrics for SCL and SCS decoding algorithms. The adjustable performance is highlighted with red lines.}
\end{figure}

From the discussions above, we can observe that SCL and SCS have complementary performance. This inspired us to combine the advantages of both and propose an adjustable decoder which can make flexible tradeoff between energy consumption and throughput for a certain given error rate performance as shown in Fig. 1. The main contributions of this work are described in the following:\par
\begin{itemize}
\item A logarithmic likelihood ratio (LLR)-threshold based path extension method is designed to reduce the memory requirements for stack decoding.\par
\item A new list-aided successive cancellation stack (LSCS) decoding algorithm is proposed. The tradeoff between its computational complexity and time complexity can be adjusted freely for a certain given error rate performance. Thus, it can meet requirements of different applications at as low computational complexity and computing resource utilization as possible.\par
\item An enhanced ELSCS algorithm is proposed on the basis of LSCS in order to further decrease the time complexity without escalating computational complexity.\end{itemize}\par
The rest of this paper is organized as follows. Section \ref{sec:pre} gives a brief introduction to the basic concept of polar codes and relevant decoding algorithms. In Section \ref{sec:llr}, an LLR-threshold based path extension method is provided, based on which we propose a novel LSCS algorithm in Section \ref{sec:lscs}. Afterwards, an enhanced version ELSCS is presented in Section \ref{sec:enhanced} while the simulation results are shown in Section \ref{sec:Numerical results}. Finally, Section \ref{sec:Conclu} gives a summary of the paper.\par

\section{Preliminaries}\label{sec:pre}

\subsection{Notation Conventions}
In this paper, we use the notation $a_i^j$ to represent the vector $\left(a_i,\;a_{i+1},\;\dots,\;a_j\right)$ and write $a_{i,e}^j$, $a_{i,o}^j$ to denote the subvector of $a_i^j$ with even and odd indices, respectively. For a positive integer $m$, $\llbracket m \rrbracket\triangleq\{1,2,...,m\}$.\par
Only square matrices are used in this paper and ${\mathbf F}_N$ represents an $N\times N$ matrix $\mathbf F$. $\left(\mathbf F\right)^{\otimes n}$ denotes the $n$-th Kronecker power of $\mathbf F$.\par
Letters $\mathcal{A}$ and $\mathcal{B}$ stand for memory stacks and $|\mathcal{A}|$ denotes the number of elements in $\mathcal{A}$. Letter $\gamma$ denotes the energy per bit to noise power spectral density ratio ($E_b/N_o$). Throughout this paper, $\ln(\cdot)$ represents the natural logarithm and $\log(\cdot)$ indicates ``logarithm to base $2$''.

\subsection{Polar Codes}
Let $W:x\rightarrow y$ denote a binary discrete memoryless channel with input alphabet $\mathbb{X}=\{0,1\}$, output alphabet $\mathbb{Y}$ and channel transition probabilities $\{W(y\vert x):\;x\in\mathbb{X},\;y\in\mathbb{Y}\}$. The channel polarization operation in Fig. 2 takes place recursively on $N$ independent copies of channel $W$. After channel polarization, the transition probability of the $i$-th synthesized subchannel is given by
    \begin{equation}
    W_N^{(i)}(y_1^N,u_1^{i-1}\vert u_i)=\sum_{u_{i+1}^N\in\chi^{N-i}}\frac1{2^{N-1}}W_N(y_1^N\vert u_1^N)
    \end{equation}

where

    \begin{equation}
    W_N(y_1^N\vert u_1^N)=\prod_{i=1}^NW(y_i\vert x_i)
    \end{equation}
\par

Although the total channel capacity remains unchanged after channel polarization, the capacity of synthesized subchannels has been polarized. It was proven by Arikan \cite{arikan2009} that as block-length approaches infinity, a portion of subchannels attains a capacity close to 1 whereas the capacity of the remaining subchannels are proximate to zero. The subchannels with higher quality are used for sending information bits, while the remaining ones are assigned fixed bits that are known to both sender and receiver in advance. This improves the reliability of useful information transmission and becomes the theoretical basis of polar codes.\par

\begin{figure}[!htbp]
\setlength{\abovecaptionskip}{1cm}

\centering

\scalebox{0.6}{\includegraphics{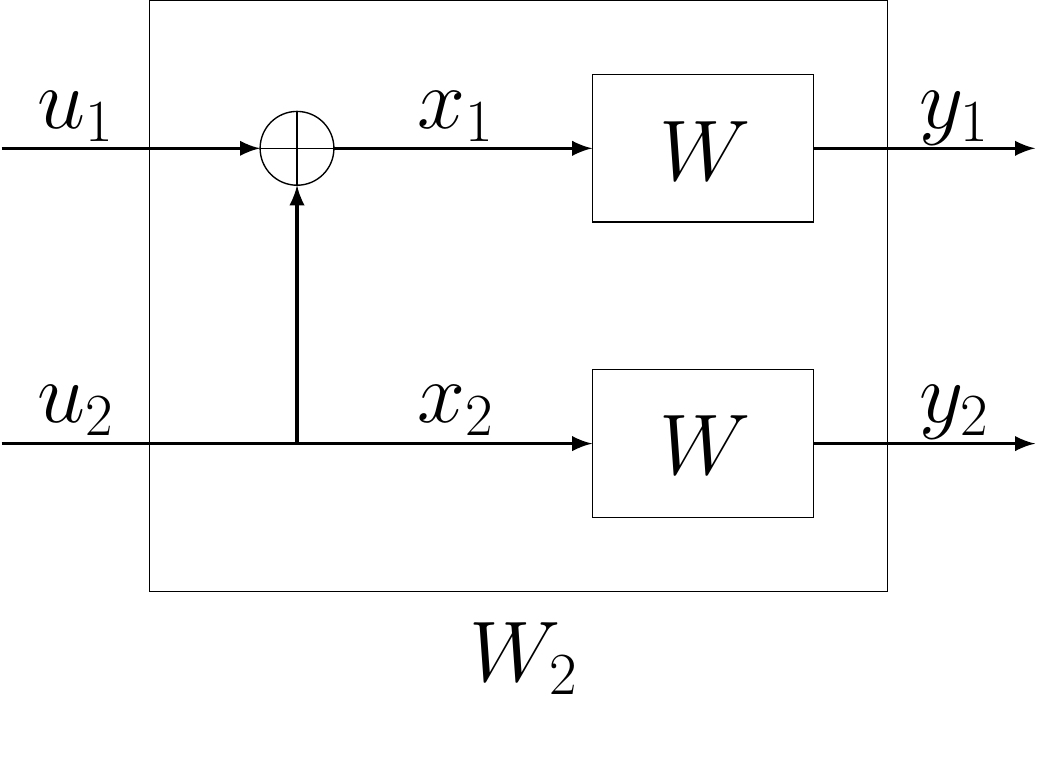}}
\caption{The channel $W_2$ and its relation to $W$.}

\end{figure}
For a polar code with block-length $N=2^n$ and rate $R$, the generator matrix can be written as ${\mathbf G}_N=B_N{\mathbf G}_2^{\otimes n}$, where ${\mathbf G}_2=\begin{bmatrix}1&0\\1&1\end{bmatrix}$, $B_N$ is an $N\times N$ bit-reversal permutation matrix. Any codeword $v_1^N$ of polar codes can be expressed as $v_1^N=u_1^N{\mathbf G}_N$, where $u_1^N$ is the information sequence consisting of $N\cdot R$ unfrozen bits and $N\cdot(1-R)$ frozen bits. The unfrozen bits correspond to the $N\cdot R$ subchannels with higher channel reliability and the frozen bits correspond to the rest \cite{tal2013construct}. Without loss of generality, all frozen bits are zero-valued.

\subsection{Decoding Algorithms}

The error rate performance of polar codes mainly depends on the decoding algorithm \cite{wu2014construction}. SC is the first proposed decoding algorithm for polar codes. SCL and SCS are derived from SC but incorporate remarkable improvements to achieve significant error rate performance gain. Moreover, when SCL and SCS are concatenated with cyclic redundancy check \cite{niu2012crc,guo2016multi}, their decoding performance can be further boosted.\par
Let the logarithmic likelihood ratio (LLR) of bit $u_i$ be defined by
    \begin{equation}
\mathrm{L}_N^{(i)}=\ln\frac{W_N^{(i)}(y_1^N,\;{\displaystyle\widehat u}_1^{i-1}\vert u_i=0)}{W_N^{(i)}(y_1^N,\;\widehat u_1^{i-1}\vert u_i=1)}
    \end{equation}
where ${\widehat u}_i$ and $y_1^N$ denote an estimate of $u_i$ and the received sequence from the channel, respectively. $\mathrm{L}_N^{(i)}$ can be calculated recursively using the equation below.
\begin{equation}
\begin{aligned}
\begin{split}
&\mathrm{L}_N^{(2i-1)}(y_1^N,\;{\widehat u}_1^{2i-2})\\&=\mathrm{L}_{N/2}^{(i)}(y_1^{N/2},\;{\widehat u}_{1,o}^{2i-2}\oplus{\widehat u}_{1,e}^{2i-2})\boxplus \mathrm{L}_{N/2}^{(i)}(y_{N/2+1}^N,{\widehat u}_{1,e}^{2i-2}\;)\;\\
\\
&\mathrm{L}_N^{(2i)}(y_1^N,\;\widehat u_1^{2i-1})=(1-{\widehat u}_{2i-1})\mathrm{L}_{N/2}^{(i)}(y_1^{N/2},\;\widehat u_{1,o}^{2i-2}\oplus\widehat u_{1,e}^{2i-2})\\&+\mathrm{L}_{N/2}^{(i)}(y_{N/2+1}^N,\;\widehat u_{1,e}^{2i-2})
\end{split}
\end{aligned}
\end{equation}
where $\boxplus$ is defined as $\ln\frac{1+e^{\alpha+\beta}}{e^\alpha+e^\beta}$ and can be approximated as a more hardware-friendly function $\mathrm{sign}(\alpha)\mathrm{sign}(\beta)\mathrm{min}\{\vert\alpha\vert,\vert\beta\vert\}$.

SC decoding employs sequential decoding. If $u_i$ is a frozen bit, its estimated value is directly set to zero, otherwise, the estimated value is given by

    \begin{equation}
\widehat u_{i}=\left\{\begin{array}{l}0\;\;\;\;\mathrm{if}\;\mathrm{L}_N^{(i)}\geq0\\1\;\;\;\;\mathrm{if}\;\mathrm{L}_N^{(i)}<0\end{array}\right.
    \end{equation}
Unlike the SC decoder, which uses hard-decision for each bit and has only one search path, SCL and SCS extend the decoding path to two new paths by appending a bit 0 or a bit 1 when an unfrozen bit is encountered. Fig. 3 and Fig. 4 show simple examples of the tree search process of SCS and SCL decoding algorithms, respectively. The bold branches represent the decoding path. The number next to each node gives the \textit{a posteriori} probability of the decoding path from the root to that node. The nodes generated after path extensions are represented by the numbered circles with the numbers indicating the extension stage order and the node with red number is the correct decoding path. The gray ones are nodes that are not visited during the search process. \par
SCL will firstly sort the new generated paths by their path metrics, which is defined by 
    \begin{equation}
{\mathrm{PM}}_i=\ln\left(P\left(\widehat u_1^i\vert y_1^N \right)\right)
    \end{equation}
${\mathrm{PM}}_i$ represents the path metric when the decoding length is $i$ whereas metric will be used as an abbreviation for path metric in the following. Thereupon, a maximum of $L$ ($L$ denotes list size) paths with the same length and the largest metrics are selected for next path extension stage without storing other paths. The decoding length is increased by one at each path extension stage. When it reaches the block-length $N$, the path with the maximum metric is output as decoding sequence. SCS stores all generated paths in a stack and selects the top path with the maximum metric to extend each time. An additional parameter, search width $Q$, is added to limit the number of extending paths with certain lengths in the decoding process. Whenever the top path with the largest metric in the stack reaches block-length $N$, the decoding process stops and outputs this path. Compared with SC, SCL and SCS have much wider search range and less probability to fall into a local optimum, which certainly ensures an improved error rate performance.\par

\begin{figure}[!htbp]
\centering
\includegraphics{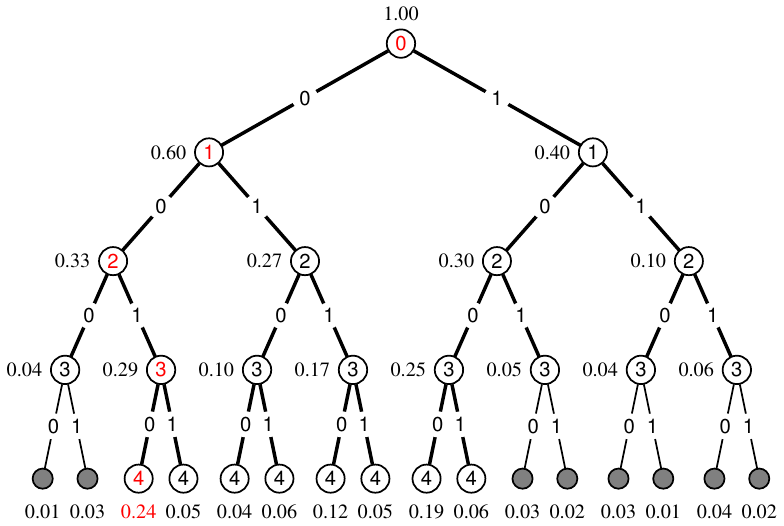}
\caption{An example of the SCL decoding with $L=4$.}
\end{figure}

\begin{figure}[!htbp]
\centering
\includegraphics{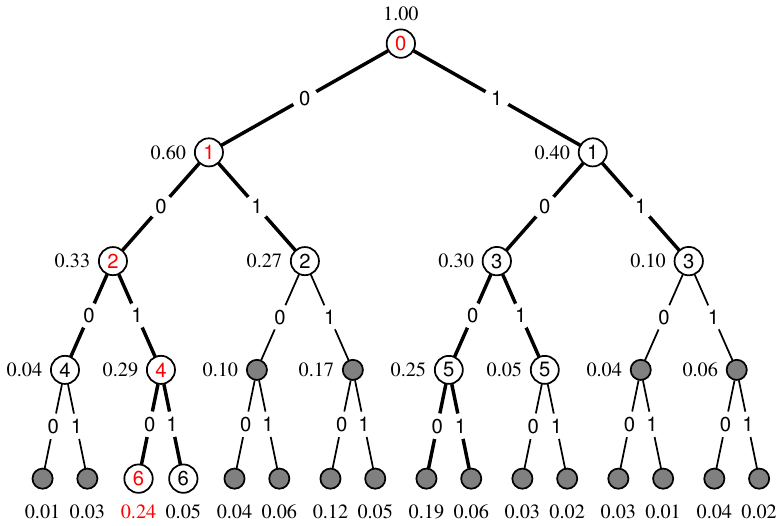}
\caption{An example of the SCS decoding.}
\end{figure}

  As SCL is a depth-first search algorithm, it extends $L$ paths with the largest metrics at the same length each time and the decoding length keeps increasing. Different from SCL, SCS employs breadth-first search, thus selecting only one path with the maximum metric to extend each time. The decoding path length in SCS is not always increasing after each extension. Therefore, it takes the decoding path a long time to reach block-length $N$. This explains why SCS has a larger decoding latency compared to SCL. As shown in the examples, SCL only needs to extend $4$ times to complete the decoding while SCS needs $2$ more. Another difference is that SCS uses a stack to store some candidate partial decoding paths with different lengths. In case of good channel quality, the correct path has a relatively larger metric than incorrect ones. For the sake of using a stack, SCS can selectively choose candidate partial path with large metric to search instead of searching $L$ paths blindly like SCL. Therefore, compared to SCL with the same error rate performance (when $Q=L$), a lot of unnecessary search is avoided. The computational complexity of SCL is approximately $L\cdot N\cdot\log N$ and that of SCS is variable and much less than that of the former. We can see from the two examples, $5$ path extension operations are saved by SCS compared to SCL.\par
  In CRC-aided decoding, the source information contains a built-in CRC and it chooses the path with the maximum metric from those who pass CRC check as output when the decoding length reaches $N$. In the case of poor channel quality, the correct path does not necessarily have the largest metric, but certainly passes the CRC check. For this reason, CRC aided decoding can further improve the probability of successful decoding.

\section{LLR-threshold based path extension scheme}
\label{sec:llr}

To combine the advantages of SCS and SCL, we need to adopt both stack decoding and list decoding. However, in stack decoding, much space is occupied by the stack to store candidate partial decoding paths with different lengths. If list decoding is further considered, more candidate partial paths need to be stored, which will occupy considerable device memory. Aiming at decreasing the required space, we propose an LLR-threshold based path extension (LTPE) scheme. Before illustrating it, a conjecture is given, which will provide theoretical basis for the scheme. If the decoding bit of the correct path is an unfrozen bit, the below phenomenon happens with a large probability. Among the two generated paths after path extension, the metric of one path remains almost unchanged and that of the other path will decrease a lot. The higher reliability of the corresponding polarization channel, the greater probability of this phenomenon. Next, we will give an empirical evidence of this conjecture.\par

The update function of metric used for path selection is
    \begin{equation}
\mathrm{PM}_{i}\lbrack \ell\rbrack=\mathrm{PM}_{i-1}\lbrack \ell\rbrack-\ln(1+e^{-(1-2{\widehat u}_i\lbrack \ell\rbrack)\cdot \mathrm{L}_N^{(i)}\lbrack \ell\rbrack})
    \end{equation}
\cite{balatsoukas2015llr}. $\mathrm{PM}_{i}\lbrack \ell\rbrack$
represents the metric of path $\ell$ with length $i$. $\widehat{u}_{i}\left [ \ell \right ]$ represents the estimated value of the $i$-th bit of path $\ell$. $\mathrm{L}_N^{(i)}\left[\ell\right]$ represents $\mathrm{L}_N^{(i)}$ of path $\ell$. The update function can be expressed as $\mathrm{PM}_{i}\lbrack \ell\rbrack=\phi(\mathrm{PM}_{i-1}\lbrack \ell\rbrack,\;\mathrm{L}_N^{(i)}\lbrack \ell\rbrack,\;{\widehat u}_i\lbrack \ell\rbrack)$, where the function $\phi(\mu,\;\lambda,\;u)\overset\triangle=\mu-\ln(1+e^{-(1-2u)\lambda})$ can be approximated to \begin{equation}
\widetilde\phi(\mu,\;\lambda,\;u)\overset\triangle=\left\{\begin{array}{l}\mu\;\;\;\;\;\;\;\;\;\;\;\mathrm{if}\;u=\frac12\lbrack1-\mathrm{sign}(\lambda)\rbrack\\\mu-\vert\lambda\vert\;\;\;\mathrm{otherwise}\end{array}\right.
\end{equation}
Thus, the update function can be approximately recast as
\begin{equation}
\widetilde{\mathrm{PM}}_i\lbrack \ell\rbrack\overset\triangle=\left\{\begin{array}{l}\mathrm{PM}_{i-1}\lbrack \ell\rbrack\;\;\;\;\;\;\;\;\;\;\;\;\;\mathrm{if}\;{\widehat u}_i\lbrack \ell\rbrack=\frac12\lbrack1-\mathrm{sign}(\mathrm{L}_N^{(i)}\lbrack \ell\rbrack)\rbrack\;\\\mathrm{PM}_{i-1}\lbrack \ell\rbrack\;-\vert \mathrm{L}_N^{(i)}\lbrack \ell\rbrack\vert\;\;\;\;\;\mathrm{otherwise}\end{array}\right.
\end{equation}

Next, we use $\mathrm{L}_i$ to represent $\mathrm{L}_N^{(i)}$ of the correct path and analyze the value of $\mathrm{L}_i$. Consider a Gaussian channel, under the assumption of Gaussian approximation, $\mathrm{L}_i\sim \mathcal{N}(\mathbb{E}\lbrack \mathrm{L}_i\rbrack,\;2\vert \mathbb{E}\lbrack \mathrm{L}_i\rbrack\vert)$ \cite{zhang2016split}, where,

    \begin{equation}
\mathbb{E}\lbrack \mathrm{L}_i\rbrack=\left\{\begin{array}{l}\mathbb{E}\lbrack \mathrm{L}_i(0)\rbrack\;\;\;\;\;\;\mathrm{if}\;u_i=0\\-\mathbb{E}\lbrack \mathrm{L}_i(0)\rbrack\;\;\;\mathrm{if}\;u_i=1\end{array}\right.
    \end{equation}
$\mathbb{E}\lbrack \mathrm{L}_i(0)\rbrack$ denotes the expectation of $\mathrm{L}_i$ for the all-zero code word transmitted over a binary-input additive white Gaussian noise channel (BI-AWGNC) with binary phase shift keying (BPSK) modulation and noise variance $\sigma_n^2$, which can be calculated recursively by the following equations.

    \begin{equation}
\begin{array}{l}m_1^{(1)}=2/\delta_n^2\\m_{2n}^{(2j-1)}=\varphi^{-1}(1-\lbrack1-\varphi(m_n^{(j)})\rbrack^2)\\m_{2n}^{(2j)}=2m_n^{(j)}\\\mathbb{E}\lbrack \mathrm{L}_i(0)\rbrack=m_N^{(i)}\end{array}
    \end{equation}
where\\
\begin{equation}
 \begin{array}{l}\varphi(x)=\left\{\begin{array}{l}1-\frac1{\sqrt{4|x|}}\int_{-\infty}^\infty \tanh\frac u2\cdot \exp(-\frac{(u-x)^2}{4|x|})du\;x\neq0\\0\;\;\;\;\;\;\;\;\;\;\;\;\;\;\;\;\;\;\;\;\;\;\;\;\;\;\;\;\;\;\;\;\;\;\;\;\;\;\;\;\;\;\;\;\;\;\;\;\;\;\;\;\;\;\;\;\;\;\;\;\;\;x=0\end{array}\right.\\\\\end{array}
\end{equation}
Given $\delta_n^2$, $\mathbb{E}\lbrack \mathrm{L}_i(0)\rbrack$ and $\mathbb{E}\lbrack \mathrm{L}_i\rbrack$ can be calculated in an off-line manner.

The relationship between the reliability of polarized channel and $\mathbb{E}\lbrack \mathrm{L}_i\rbrack$ is shown as below.

    \begin{equation}
\begin{array}{l}P_e(u_i)=\int_{-\infty}^0\frac1{2\sqrt{\mathrm\pi |\mathbb{E}\lbrack \mathrm{L}_i\rbrack|}}\cdot \exp\left(\frac{-(x-|\mathbb{E}\lbrack \mathrm{L}_i\rbrack|)^2}{4|\mathbb{E}\lbrack \mathrm{L}_i\rbrack|}\right)dx\\\;\;\;\;\;\;\;\;\;=Q(\sqrt{|\mathbb{E}\lbrack \mathrm{L}_i\rbrack|/2})\\\\\end{array}
    \end{equation}
where $P_e(u_i)$ represents the probability that $u_i$ is incorrectly estimated in terms of the subchannel $W_N^{(i)}(y_1^N,\;u_1^{i-1}\vert u_i)$, given the correct prior bits $u_1^{i-1}$. The smaller $P_e(u_i)$, the higher the channel reliability. $Q(x)=\frac1{\sqrt{2\mathrm\pi}}\int_x^{+\infty}e^{-\frac{t^2}2}\operatorname dt$ is a monotonically decreasing function, thus the channel reliability increases with the increasing value of $|\mathbb{E}\lbrack \mathrm{L}_i\rbrack|$.\par

The unfrozen bits are chosen according to the channel reliability and corresponds to the $N\cdot R$ polarization channels with the smallest $P_e(u_i)$ and the largest $|\mathbb{E}\lbrack \mathrm{L}_i\rbrack|$. Consequently, for unfrozen bit $i$,
$|\mathrm{L}_i|$ has a high probability to be a large value, and the greater reliability  polarization channel $i$ has, with higher probability this happens. We can see from Fig. 5, where the points above 40 are projected onto the $\geq40$ line for convenience, that when $N=1024$, $\gamma=2$ dB, the smallest $|\mathbb{E}\lbrack \mathrm{L}_i\rbrack|$ for unfrozen bits is $14.2$, the values at some other $\gamma$ are also shown in Table 1.

\begin{figure}[!htbp]
\setlength{\abovecaptionskip}{0.cm}
\flushleft
\includegraphics{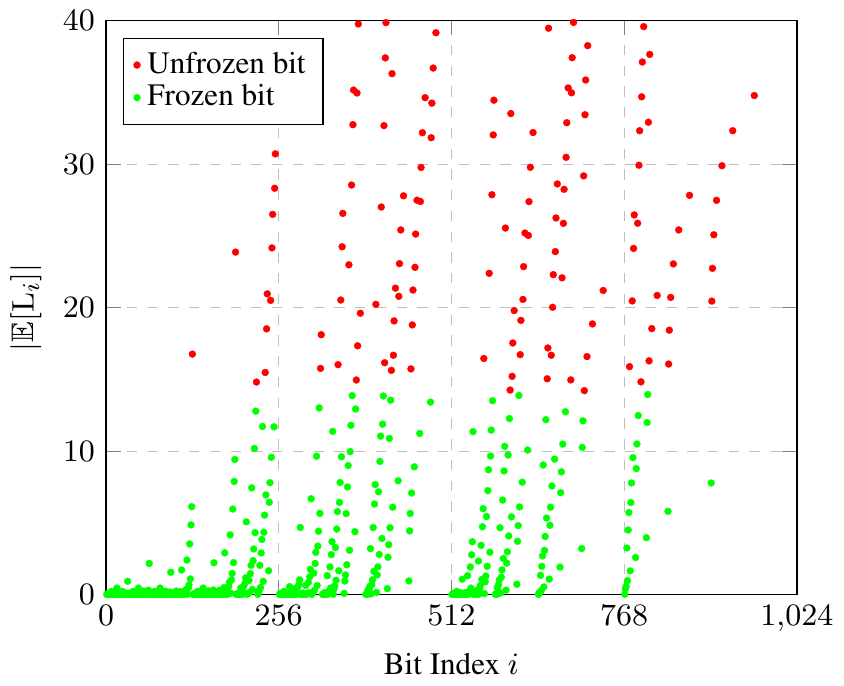}
  \begin{picture}(0,0)  
\put(10,211){\small $\geq$} 
  \end{picture} 
\caption{$|\mathbb{E}\lbrack \mathrm{L}_i\rbrack|$ with different bit indices $i$.}
\end{figure}

\begin{table}[H]
\caption{$\mathrm{Min}(|\mathbb{E}\lbrack \mathrm{L}_i\rbrack|)$ for unfrozen bits with different $\gamma$}
\centering
\begin{tabular}{|c|c|c|c|c|c|}
\hline
$\gamma$ ($N=1024$) & $1.0$ dB & $1.5$ dB & $2.0$ dB & $2.5$ dB & $3.0$ dB\\
\hline
\makecell{$\min(|\mathbb{E}\lbrack \mathrm{L}_i\rbrack|)$,\\$i$ is unfrozen bit} & $5.38$ & $9.38$ & $14.20$ & $23.00$ & $30.00$\\
\hline
\end{tabular}
\end{table}
When the correct decoding path comes to unfrozen bit $i$ in a sequential decoding, the value of $|\mathrm{L}_i|$ is large, especially when the corresponding polarization channel has high reliability. Consider Eq. (9), it happens with large probability that the metric of one path remains almost unchanged after path extension and another will decrease considerably. \par
The one with smaller metric is very difficult to become the final decoding because of the big metric gap. However, these kind of paths occupy a lot of storage space. Hence, we consider to set an LLR threshold. If the metric difference is larger than the threshold, the one with smaller metric will be deleted. The other one whose metric almost remains constant will continue to be extended at the next path extension stage and does not need to be considered in sorting stage. Let $\delta\;(\delta>0)$ denotes the LLR threshold. The LTPE scheme is summarized in Algorithm $1$.

\begin{algorithm}
$\mathcal{A}$ denotes the stack for storing paths to be extended\;
$\mathcal{B}$ denotes the stack for storing generated paths after extension\;
\eIf{$u_{i+1}$ is unfrozenbit}{\eIf{$\vert \mathrm{L}_{i+1}\lbrack \ell\rbrack\vert\geq\delta$}
{Delete $(u_1^i\lbrack \ell\rbrack,\;u_{i+1})$, $u_{i+1}\neq\frac12\lbrack1-\mathrm{sign}(\mathrm{L}_{i+1}\lbrack \ell\rbrack)\rbrack$\;
Reserve $(u_1^i\lbrack \ell\rbrack,\;u_{i+1})$, $u_{i+1}=\frac12\lbrack1-\mathrm{sign}(\mathrm{L}_{i+1}\lbrack \ell\rbrack)\rbrack$ in stack $\mathcal{A}$ for next path extension stage directly\;}
{Push both paths into stack $\mathcal{B}$ for sorting process\;}}
{Push $(u_1^i\lbrack \ell\rbrack,\;0)$ into stack $\mathcal{B}$ for sorting process\;}
\caption{LLR-threshold Based Path Extension Scheme}
\end{algorithm} 

In Algorithm 1, $\mathrm{L}_{i+1}\lbrack \ell \rbrack$ represents $\mathrm{L}_N^{(i+1)}$ of the decoding path $\ell$, $(u_1^i\lbrack \ell \rbrack,\;u_{i+1})$ denotes the generated new path after path $\ell$ chooses $u_{i+1}$ to extend at the $(i+1)$-th bit. This LTPE strategy can reduce the number of paths inserted to the stack, avoiding the excessive space occupied by those paths that are difficult to become the final decoding output. As fewer bits are required for the quantization of LLR than metric in hardware implementation, it is easier to compare LLR than metric \cite{balatsoukas2015llr}. In addition, the deleted paths in LLR-based scheme are not required to execute Eq. (9) for metric computation. Hence, the proposed LLR-based path pruning scheme is more convenient to implement in practice, compared with existing metric-based path pruning schemes. Of course, we can also use metric-based schemes after the proposed LLR-based scheme to further lower the memory requirements, at the cost of higher implementation complexity.\par
For how to suitably select the threshold $\delta$ to reduce the storage requirements while maintaining error rate performance unaffected, we will give an in-depth analysis and discussion in simulation part in Section \ref{sec:Numerical results}.

\section{List-aided successive cancellation stack decoding algorithm}
\label{sec:lscs}

Motivated by the advantage of SCL, we introduce list decoding to speed up the breadth search of stack decoding. A new LSCS decoding is proposed on the basis of the path extension method in Algorithm $1$. At each extension stage, LSCS chooses a maximum of $L$ paths with the largest metrics in the stack to extend simultaneously. The detailed steps are shown in Algorithm $2$ and an example is given in Fig. 6. This algorithm can offer a flexible tradeoff between time complexity and computational complexity by adjusting the value of $L$. The adjustable complexity performance varies among maximum time/minimum computational complexity (similar to SCS) and minimum time/maximum computational complexity (similar to SCL) and some intermediate complexity states freely, while ensuring a stable error rate performance.\par

\begin{algorithm}
Define $Q$ as the maximum tolerable number of CRC check\;Define $L\;(1\leq L\leq Q)$ as the number of paths that are extended simultaneously at each extension stage\;
\textbf{1) Initialization}\\
Create a stack $\mathcal{A}$ with depth $L$\;
Create a stack $\mathcal{B}$ with depth $D$\;
Push a null path into $\mathcal{A}$, set path metric to 0\;
Initialize counter, $q_1^N=zeros(N,1)$\;
\textbf{2) Path competition}\\
Define $u_1^i\lbrack \ell\rbrack$ as path $\ell$ with a length $i$ in stack $\mathcal{A}$\;
\If{$u_1^i\lbrack \ell\rbrack\neq null$}{$q_i=q_i+1$\; \If{$q_i=Q$}
{Delete all paths in $\mathcal{B}$ with length less than or equal to $i$\;}}
\textbf{3) Path extension}\\
Extend the paths in stack $\mathcal{A}$ according to Algorithm 1\;
\textbf{4) Path selection and sorting}\\

After all $L$ paths in stack $\mathcal{A}$ complete path extension\;
\eIf{$|\mathcal{A}|=L$}{
Sort the paths in stack $\mathcal{B}$ in descending metric\;
}{
Push the $L-|\mathcal{A}|$ paths with the maximum metrics in stack $\mathcal{B}$ to stack $\mathcal{A}$ and sort the rest paths in stack $\mathcal{B}$ in descending metric\;
}
\textbf{5) CRC-aided termination decision}\\

\eIf{exist paths in $\mathcal{A}$ with a length $N$}{
Perform CRC detection\;

\eIf{CRC detection pass}{Output the path with the maximum metric as decoding sequence\;}
{$q_N=q_N+1$\; \eIf{$q_N<Q$}{Go to step 2\;}{Declare a decoding failure\;}}}{Go to step 2\;}

\caption{The LSCS($Q$, $L$, $D$) Decoder}
\end{algorithm} 

\begin{figure}[!htbp]
\centering
\includegraphics{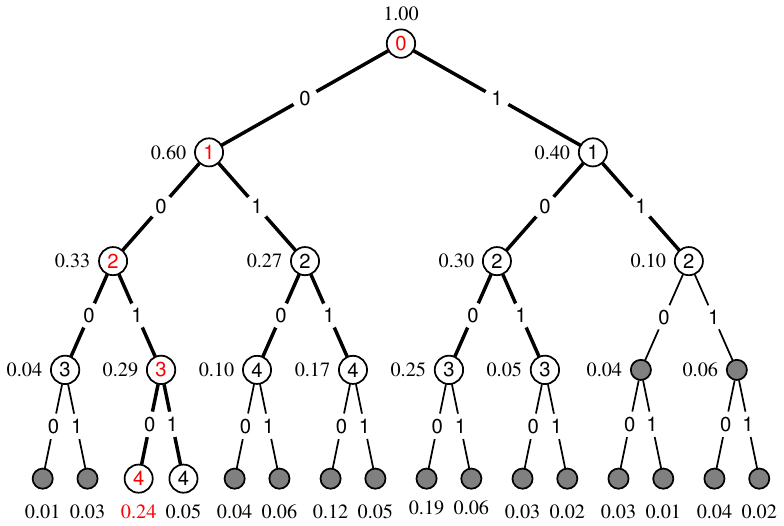}
\caption{An example of the LSCS decoding with $L=2$, $Q=4$.}
\end{figure}

Considering the iterative property of LLR operations in Eq. (4), it is difficult to be completed synchronously, which directly affects the decoding delay and throughput. For example, ${\log}N$ iterations are required to calculate the value of $\mathrm{L}_N^{(1)}$. Each iteration needs to call Eq. (4). If one clock cycle (CLK) is required for each execution of Eq. (4), the calculation of $\mathrm{L}_N^{(1)}$ requires at least ${\log}N$ CLKs. In this manner, when $L$ paths are processed in parallel, to simplify the time complexity evaluation of LSCS decoding,  we can suppose that all the parallelizable executions of Eq. (4) are performed in one CLK \cite{hashemi2016fast} and measure the time complexity in terms of the maximum number of CLKs for each decoding.\par  
The time complexity of LSCS decoding is given by
    
    \begin{equation}
{\mathrm T}_\mathrm{LSCS}\triangleq\sum_{k=1}^K\left(\underset{j\in{\llbracket L \rrbracket}}{\mathrm{max}}{\mathcal {O}}_\mathrm t\left({\ell}_\mathit j^\mathit {\left(k\right)}\right)\right)
    \end{equation}    
where ${\ell}_\mathit j^\mathit {(k)}$ denotes the $j$-th extending path at the $k$-th extension stage. ${\mathcal {O}}_\mathrm t(\cdot)$ is the time complexity calculation operation and ${\mathcal {O}}_\mathrm t({\ell}_\mathit j^\mathit {(k)})$ represents the number of CLKs required by the $j$-th path at the $k$-th extension stage, which is in the range of $1$ to ${\log}N$. $K$ denotes the total number of extension stages in the decoding process.\par

For the computational complexity, we refer to \cite{niu2012crc} and use the number of LLR operations in Eq. (4) to measure the decoding computational complexity, which determines the amount of energy consumption approximately. The computational complexity of LSCS decoding is expressed as
    \begin{equation}
{\mathrm C}_\mathrm{LSCS}\triangleq\sum_{k=1}^K\left(\sum_{j=1}^L\left({\mathcal {O}}_\mathrm c\left({\ell}_\mathit j^\mathit {(k)}\right)\right)\right)
    \end{equation}
where ${\mathcal {O}}_\mathrm c(\cdot)$ is the computational complexity calculation operation. ${\mathcal {O}}_\mathrm c({\ell}_j^{(k)})$ represents the number of LLR operations required by the $j$-th path at the $k$-th extension stage, which is in the range of 1 to ${\log}N$.\par

Because the change of list size only influences the search speed and is unrelated with the search range, the error rate performance will not be affected. It is expected that with the increasing number of paths searched simultaneously at each extension stage, the search speed accelerates and LSCS will need less extension stages $K$ to find the correct path compared with SCS. Therefore, according to Eq. (14), this kind of parallelized search strategy will reduce the average decoding time complexity. However, the downside is the increase in computational complexity. Although $K$ decreases, the reduction magnitude of $K$ is not proportional to the size of $L$. This is because more incorrect paths will be searched when $L$ increases, which reduces the search efficiency. Meanwhile, $\sum_{j=1}^L{\mathcal {O}}_\mathrm c({\ell}_\mathit j^\mathit {(k)})$ increases almost linearly with the size of $L$. The increasing number of LLR operations for an extension stage becomes the dominant influencing factor instead of the decreasing $K$. As a consequence, according to Eq. (15), the average computational complexity will increase.\par 
When $L=1$, the LSCS is similar to SCS, extending one path each time. In this case, the time complexity is maximized whereas the computational complexity is minimized. When $L=Q$, LSCS is similar to SCL, extending $Q$ paths with the same length simultaneously, the time complexity is minimized whereas the computational complexity is maximized at this time. When $L$ increases from $1$ to $Q$, the time complexity decreases and the computational complexity increases. Later numerical simulation results match well with the above analysis. LSCS can flexibly change its complexity performance among the two extreme states and their intermediate states by adjusting the list size while the error rate performance remains unaltered.

\section{Enhanced list-aided successive cancellation stack decoding algorithm}
\label{sec:enhanced}

Conversely to SCL, where all the $L$ extending paths have the same length,  in LSCS the $L$ extending paths may have different lengths if $1<L<Q$. If the length $i$ of a extending path is odd, the number of CLKs to obtain $\mathrm{L}_N^{(i+1)}$ is only $1$. Otherwise, more CLKs are required to calculate the value of $\mathrm{L}_N^{(i+1)}$. For example, it will take $\log N$ CLKs to calculate the value of $\mathrm{L}_N^{(1)}$ if $i=0$. Considering parallel processing, every extending path is assigned a processing element (PE). According to the above analysis, at the extension stage, some PEs will run for a long time while some will complete their tasks quickly and wait until other PEs stop working. This not only can not make full use of PE resources but also increases the decoding latency. To solve this problem, we allow each path extend two sequential bits at the extension stage, which means that each extension stage consists of two extensions. For example, when path $\ell$ generates new paths after the extension $1$ at its first bit $i$, it does not need to wait until all other paths finish the extension $1$ at their first bit, the new generated path with larger metric is chosen to continue to extend at the second bit $i+1$. When all paths have finished extension 2 at their second bit, the sorting operation is performed. As a result of extending two sequential bits, the indices of the two extended bits must have an odd and an even number, which makes the runtime of PEs corresponding to different paths slightly differ from each other. This improves the PE resource utilization efficiency. Thus, an enhanced LSCS (ELSCS) is proposed in Algorithm $3$ (Its steps $1$, $2$, $4$, $6$, $7$ are the same as Algorithm $2$) to further decrease the time complexity and a simple example is shown in Fig. 7.\par

\begin{algorithm}
Define $Q$ as the maximum tolerable number of CRC check\;Define $L\;(1\leq L\leq Q)$ as the number of paths that are extended simultaneously at each extension stage\;
\textbf{1) Initialization}\\
\textbf{2) Path competition}\\
\textbf{3) Path extension 1}: Extend path $u_1^i\lbrack \ell\rbrack$ to length $i+1$\\
\eIf{$u_{i+1}$ is unfrozen bit}{Reserve $(u_1^i\lbrack \ell\rbrack,\;u_{i+1})$, $u_{i+1}=\frac12\lbrack1-\mathrm{sign}(\mathrm{L}_{i+1}\lbrack \ell\rbrack)\rbrack$ in stack $\mathcal{A}$ for path extension 2 directly\;\eIf{$\vert \mathrm{L}_{i+1}\lbrack \ell\rbrack\vert\geq\delta$}{Delete another path\;}{Push another path into stack $\mathcal{B}$ for sorting process\;}}{Reserve $(u_1^i\lbrack \ell\rbrack,0)$ in stack $\mathcal{A}$ for path extension 2 directly\;}
\textbf{4) Path competition}\\
\textbf{5) Path extension 2}: Extend path $u_1^{i+1}\lbrack \ell\rbrack$ to length $i+2$\\ Extend the paths in stack $\mathcal{A}$ according to Algorithm 1\;
\textbf{6) Path selection and sorting}\\
\textbf{7) CRC-aided termination decision}\\

\caption{The Enhanced LSCS($Q$, $L$, $D$) Decoder}
\end{algorithm} 

\begin{figure}[!htbp]
\centering
\includegraphics{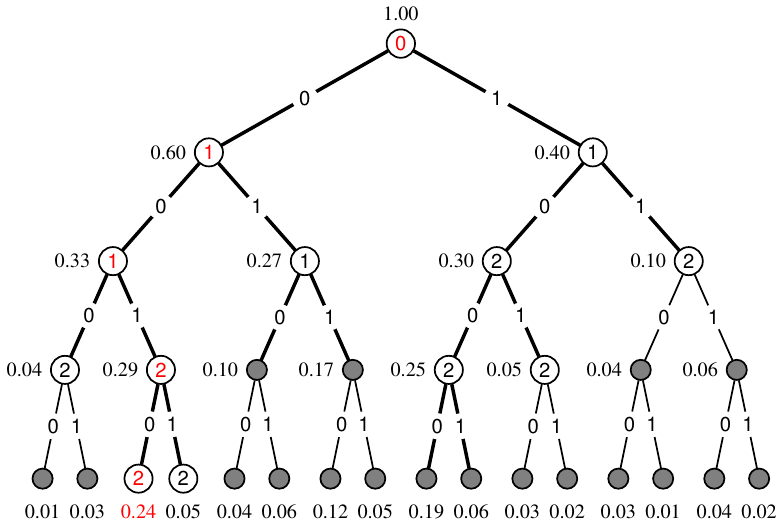}
\caption{An example of the ELSCS decoding with $L=2$, $Q=4$.}
\end{figure}

Despite existing a difference between the path extension schemes of ELSCS and LSCS, the path search range of ELSCS does not diminish compared with that of LSCS due to the use of stack. So the error rate performance will remain unchanged. \par

The time complexity of ELSCS decoding is given by

    \begin{equation}
{\mathrm T}_\mathrm{ELSCS}\triangleq\sum_{k=1}^K\left(\underset{i\in{\llbracket 2 \rrbracket},j\in{\llbracket L \rrbracket}}{\mathrm{max}}{\mathcal {O}}_\mathrm t\left({\ell}_\mathit j^\mathit {\left(k_i\right)}\right)+1\right)
    \end{equation}
where ${\ell}_\mathit j^\mathit {(k_i)}$ denotes the $j$-th extending path at extension $i$ of the $k$-th extension stage. ${\mathcal {O}}_\mathrm t({\ell}_\mathit j^\mathit {(k_i)})$ indicates the number of CLKs required by the $j$-th extending path $\ell$ at extension $i$ of the $k$-th extension stage. $\underset{i\in{\llbracket 2 \rrbracket},j\in{\llbracket L \rrbracket}}{\mathrm{max}}{\mathcal {O}}_\mathrm t({\ell}_\mathit j^\mathit {(k_i)})$ is the maximum number of CLKs required to extend paths with an even length. $1$ is the number of CLKs required to extend the paths with an odd length. $K$ denotes the total number of extension stages in the decoding process. 

The maximum number of CLKs for an extension stage $k$ in LSCS is

    \begin{equation}
\underset{j\in{\llbracket L \rrbracket}}{\mathrm{max}}{\mathcal {O}}_\mathrm t({\ell}_\mathit j^\mathit {(k)})
    \end{equation}

The maximum number of CLKs for an extension stage $k$ in ELSCS is
    \begin{equation}
\underset{i\in{\llbracket 2 \rrbracket},j\in{\llbracket L \rrbracket}}{\mathrm{max}}{\mathcal {O}}_\mathrm t({\ell}_\mathit j^\mathit {(k_i)})+1
    \end{equation}
Comparing Eq. (17) and Eq. (18), the average maximum number of CLKs for an extension stage in ELSCS may be slightly larger than that in LSCS.

However, because of extending two bits in each extension stage, the average number of extension stages required for each decoding in ELSCS are much fewer than that in LSCS. Thus, when the average total maximum number of CLKs for a decoding process is considered, ELSCS performs better than LSCS. In other words, extension $1$ reserves the extended path with larger metric for extension $2$ directly, allowing each path to extend two bits in succession without waiting for other paths, increasing the PE resource utilization efficiency. Therefore, it will take ELSCS less time to complete the computations for a decoding process compared with LSCS. \par

The computational complexity of ELSCS decoding is expressed as 
    \begin{equation}
{\mathrm C}_\mathrm{ELSCS}\triangleq\sum_{k=1}^K\left(\sum_{j=1}^L\left(\sum_{i=1}^2\left({\mathcal {O}}_\mathrm c\left({\ell}_\mathit j^\mathit {(k_i)}\right)\right)\right)\right)
    \end{equation}
where ${\mathcal {O}}_\mathrm c({\ell}_\mathit j^\mathit {(k_i)})$ indicates the number of LLR operations required by the $j$-th extending path $\ell$ at extension $i$ of the $k$-th extension stage. $\sum_{j=1}^L(\sum_{i=1}^2({\mathcal {O}}_\mathrm c({\ell}_\mathit j^\mathit {(k_i)})))$ means the number of LLR operations for an extension stage.\par

For extending two sequential bits at each extension stage, compared to LSCS, the average number of LLR operations for an extension stage in ELSCS doubles while the average number of extension stages required for each decoding in ELSCS nearly halves. Thus, the average total number of LLR operations for a decoding process in ELSCS will not differ much from that in LSCS. In Section \ref{sec:Numerical results}, simulation results also show that ELSCS has almost the same computational complexity with LSCS.\par
In terms of hardware resources, ELSCS and LSCS both need up to $Q$ PEs. The number is identical with that of SCL when they have the same error rate performance. However, the $Q$ PEs in LSCS/ELSCS are not necessarily all used. The exact number of used PEs is equal to list size $L$. If $L<Q$, only partial PEs are used for decoding, the remaining PEs can be used by other modules such as demodulation and equalization. So LSCS/ELSCS will have a lower computing resource occupancy compared with SCL at this time. Another difference is more storage space for LSCS/ELSCS when $L<Q$. As the algorithm is adjustable, if there is not enough memory, $L$ can be set equal to $Q$ in LSCS/ELSCS and the required stack depth will reduce to a minimum of $2L$ like in SCL, at the cost of higher computational complexity.

\section{Numerical results}
\label{sec:Numerical results}

Numerical simulation results for BI-AWGNC are presented in this section to compare the block error rate (BLER) and complexity of different polar codes decoding algorithms. $10^7$ code blocks are transmitted. All the used codes have code length $N=1024$ and code rate $R=1/2$, a CRC-24 code with generator polynomial $g(D)=D^{24}+D^{23}+D^6+D^5+D+1$ is used.\par

\begin{figure}[!htbp]
\setlength{\abovecaptionskip}{0.cm}
\setlength{\belowcaptionskip}{-0.cm}
\centering
\includegraphics{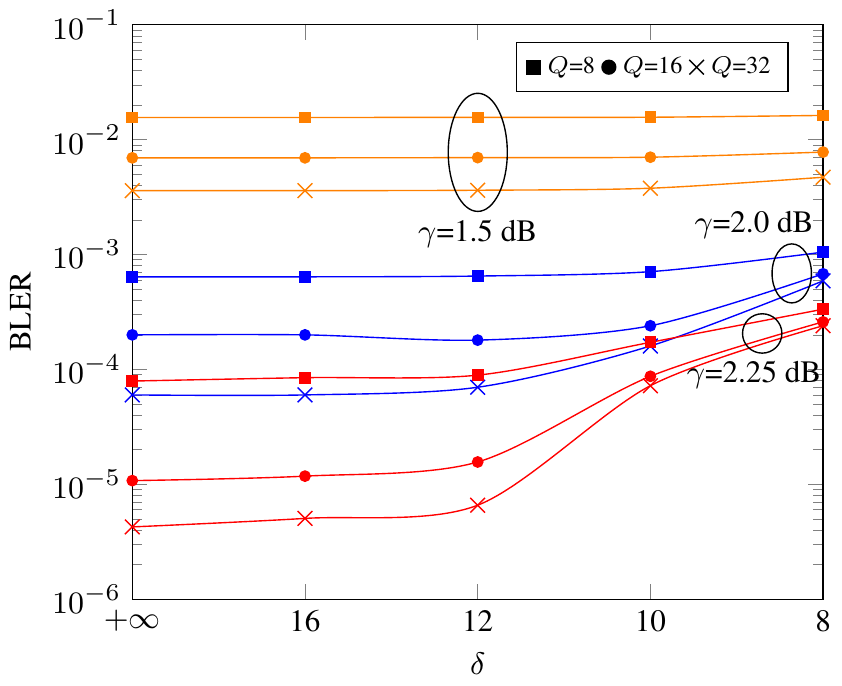}
\caption{BLER performance of ELSCS ($L$=1) with different $\delta$.}
\end{figure}

Fig. 8 shows the BLER of ELSCS decoder with different $Q$ and $\gamma$ under different LLR threshold $\delta$. The points below $10^{-6}$ are projected onto the x-axis for convenience. When $\delta$ decreases, the error rate performance degrades, especially when $\delta<12$. The explanation for this behaviour is that the correct path has a higher probability to be pruned under a smaller $\delta$, resulting in a larger decoding failure probability. It can also be observed that this trend is not significant in the low SNR regime. Because low SNR leads to small $|\mathbb{E}\lbrack \mathrm{L}_i\rbrack|$, less pruning will happen and the correct path is more likely to be retained. The BLER curves of ELSCS with different list sizes all follow the same change trend as in Fig. 8. By comprehensive consideration of both the pruning capability and influence on BLER performance, $\delta$ is set to $12$ in the following simulation analysis. The deterioration of the error rate performance is negligible at this time and a uniform integer value threshold makes threshold comparison simple and easy for hardware implementation. In addition, we find that when the value of $Q$ is $16$, an ideal error rate performance ($\mathrm{BLER}\leq2\times10^{-4}$) can be obtained from $\gamma$ greater than $2.0$ dB. When $Q=32$, although the error rate performance is superior, the computational complexity will be much higher as it is generally proportional to the value of $Q$. Thus, considering the convenience of practical realization, the value of $Q$ is set to $16$.\par

\begin{figure}[!htbp]
\setlength{\abovecaptionskip}{0.cm}
\setlength{\belowcaptionskip}{-0.cm}
\centering
\includegraphics{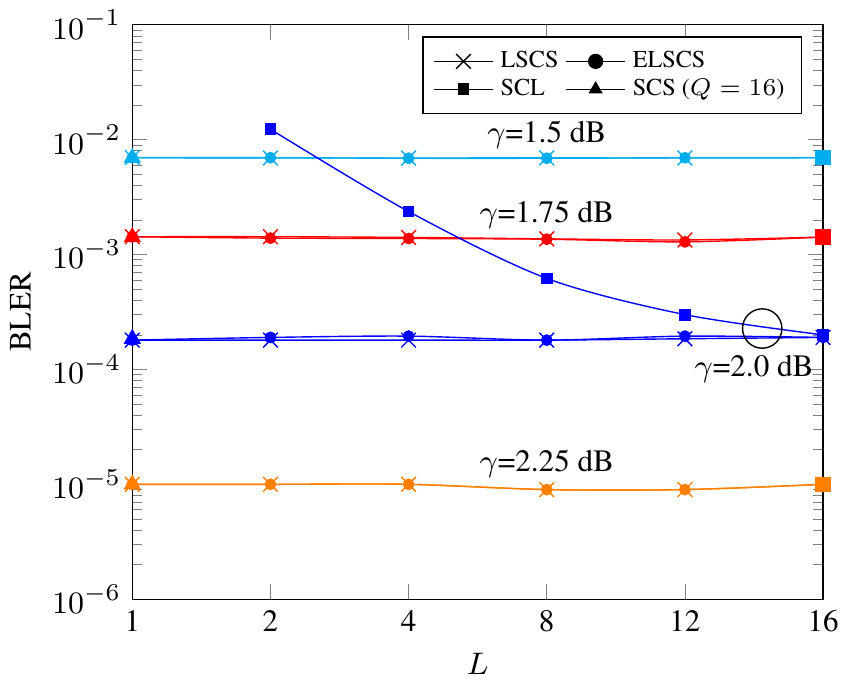}
\setlength{\abovecaptionskip}{0.cm}
\setlength{\belowcaptionskip}{-0.cm}
\caption{BLER performance of different decoding algorithms with different $L$ values and $\gamma$ when $Q=16$, $\delta=12$.}
\end{figure}

Fig. 9 illustrates that the error rate performance of LSCS/ELSCS almost remains unchanged with different $L$ values and is consistent with that of SCS and SCL. Since the increasing list size only speeds up the path search and does not reduce the path search range, the probability of successful decoding does not decline. Particularly, when $L=16$, the decoding principle and performance of ELSCS is similar to SCL, they both extend $16$ paths with the same length each time, the difference is that SCL chooses $16$ paths with the maximum metrics after sorting for next extension stage, while, ELSCS may reserve some paths for next extension directly without sorting operation. However, simulation results show that the performance deterioration caused by this difference can be neglected. This is because the path extension characteristics of the reserved paths match well with that of the correct one when $\delta$ is large enough, so this path reservation strategy barely reduces the existence probability of the correct path among the decoding paths.\par
The BLER curve of SCL with different $L$ at $\gamma=2.0$ dB is also plotted in Fig. 9. It increases with the decrease of $L$ and is higher than that of LSCS/ELSCS with the same $L$. When $L=4$, ELSCS can obtain a performance gain of  more than $0.25$ dB compared with SCL at the BLER of $1.4\times10^{-3}$ and when $L=2$, the performance gain increases to $0.5$ dB at the BLER of $7\times10^{-3}$.
\par

\begin{figure}[!htbp]
\setlength{\abovecaptionskip}{0.cm}
\setlength{\belowcaptionskip}{-0.cm}
\centering
\includegraphics{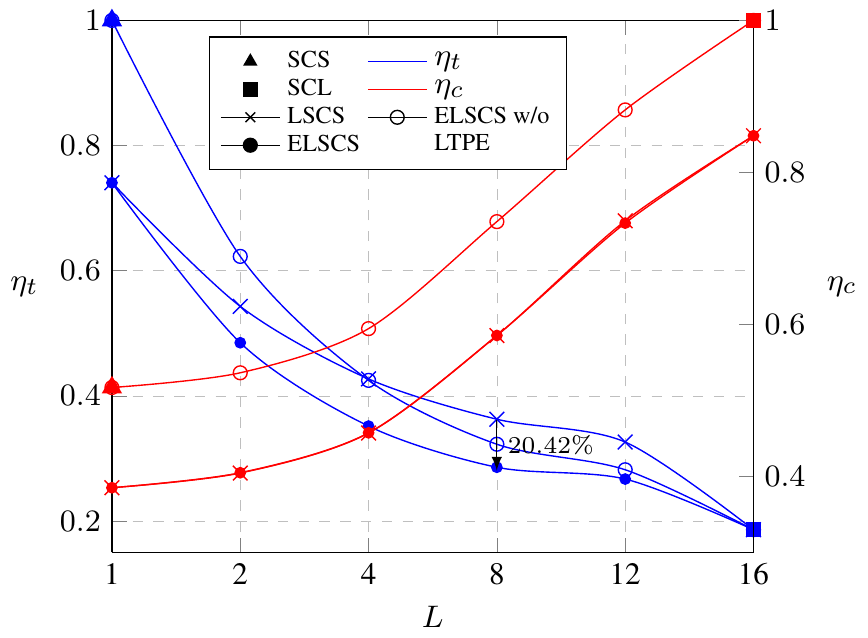}
  \begin{picture}(0,0)
  \end{picture}
\caption{Normalized time and computational complexities of different decoding algorithms with different $L$ values at $\gamma=2$ dB when $Q=16$, $\delta=12$.}
\end{figure}

\begin{figure}[!htbp]
\setlength{\abovecaptionskip}{0.cm}
\setlength{\belowcaptionskip}{-0.cm}
\centering
\includegraphics{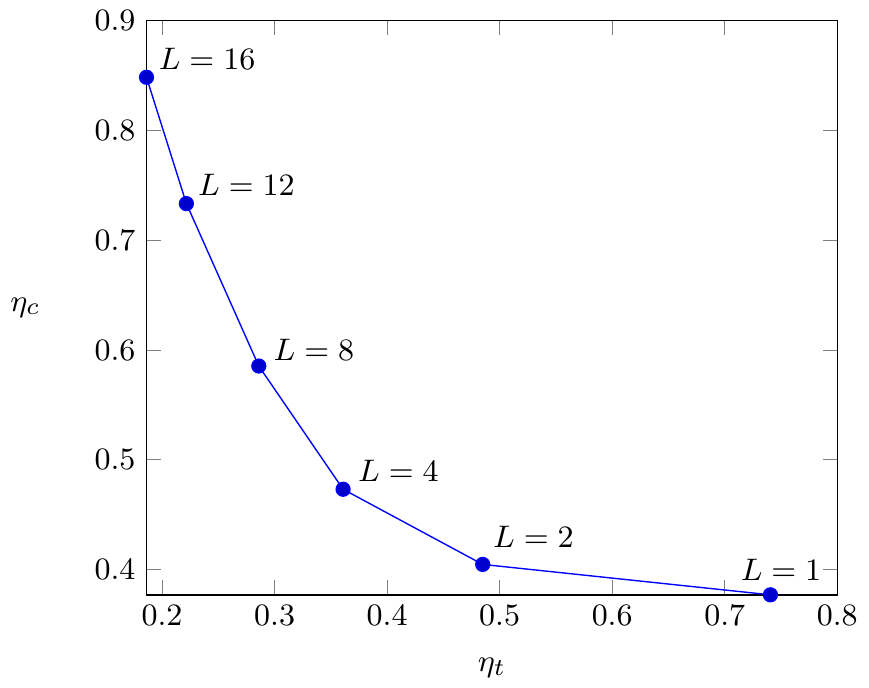}
\caption{Normalized time and computational complexities of ELSCS with different modes at $\gamma=2$ dB when $Q=16$, $\delta=12$.}
\end{figure}

Fig. 10 shows the normalized time complexity $\eta_t$ and computational complexity $\eta_c$ of different decoding algorithms with different $L$ values. They are defined by
    \begin{equation}
\eta_t=\frac{{\overline{\mathrm T}}_\ast}{{\overline{\mathrm T}}_\mathrm{SCS}}
    \end{equation} 
    \begin{equation}
\eta_c=\frac{{\overline{\mathrm C}}_\ast}{{\overline{\mathrm C}}_\mathrm{SCL}}
    \end{equation}     
where ${\overline{\mathrm T}}_\ast$ and ${\overline{\mathrm C}}_\ast$ indicate the average time and computational complexity of different decoding algorithms, respectively. ${\overline{\mathrm T}}_\mathrm{SCS}$ denotes the average time complexity of SCS with $Q=16$ and ${\overline{\mathrm C}}_\mathrm{SCL}$ denotes the average computational complexity of SCL with $L=16$. The average time and computational complexity of LSCS and ELSCS are calculated using Eq. (14), (15), (16), (19), respectively.

We can see that it is difficult to obtain both optimal computational and time complexity performance under certain given error performance. Minimum computational complexity often means maximum time complexity and vice versa. For LSCS and ELSCS, we can adjust the tradeoff between computational and time complexity by changing the list size. In Fig. 10, the BLER is constant at $2\times10^{-4}$. When $L=1$, the decoding principle and performance of ELSCS are similar to SCS, achieving the minimum computational complexity and the maximum time complexity. On the contrary, when $L=16$, ELSCS extends $16$ paths with the same length each time, similar to SCL, thus, has the maximum computational complexity and the minimum time complexity. When list size is between $1$ and $16$, there will be many intermediate performance states.
As list size increases, the average maximum number of CLKs for each extension stage may become slightly larger. However, the average number of path extension stages required for each decoding drops, leading to the declination of the average total maximum number of CLKs for each decoding. Consequently, the time complexity decreases. When considering the computational complexity, the average number of LLR operations for each extension stage increases dramatically with increasing list size and becomes the dominant influencing factor. Hence, the average total number of LLR operations rises and the computational complexity increases.\par

As a comparison, ELSCS without LTPE scheme is depicted. It can be found that LTPE scheme reduces both time and computational complexity. This is because fewer candidate partial decoding paths are inserted into the stack as a result of the path pruning capability of LTPE scheme. The decoder does not need to search among many paths that are improbable to become the final decoding output. Consequently, ELSCS with LTPE scheme can get lower decoding time and computations compared to ELSCS without LTPE scheme. In addition, ELSCS without LTPE scheme can still provide a flexible tradeoff between time complexity and computational complexity. The complexity performance of two extreme cases when $L=1$ and $L=16$ is the same as that of SCS and SCL, respectively. Thus, even without LTPE scheme, ELSCS is still able to provide a low time complexity or computational complexity through the adjustment of list size $L$.\par

We can also observe that the computational complexity curves of LSCS and ELSCS overlap but the time complexity of ELSCS reduces in comparison to that of LSCS. As ELSCS extends two sequential bits at each path extension stage, the PE resource utilization efficiency is improved and the time complexity has a significant reduction. When $L=8$, the relative time complexity of ELSCS is reduced by $20.42\%$ compared with that of LSCS.\par
As shown in Fig. 11, we can consider choosing $L=1, 2, 4, 8, 12, 16$ as different modes of ELSCS algorithm, thereby providing an adjustable decoding algorithm for different cases. A proper configuration of $L$ can make the algorithm meet the specific decoding latency and throughput demand, while, reducing the computational complexity and the occupied number of PEs to the least possible degree. As a comparison, although SCL ($L=16$) can provide smaller latency, it requires more computational complexity
and occupies more computing resources. When the decoding latency of ELSCS already meets the throughput requirement of application, a smaller decoding latency with higher cost is not necessary. In the practical application of ELSCS algorithm, the value of $Q$ is chosen firstly, which determines the error rate performance. Generally, Q is set to 16. Then the value of $L$ is adjusted to select the appropriate algorithm mode, ensuring that the required throughput is achieved with the lowest computational complexity and minimum computing resources.\par

\begin{figure}[!htbp]
\centering
\includegraphics{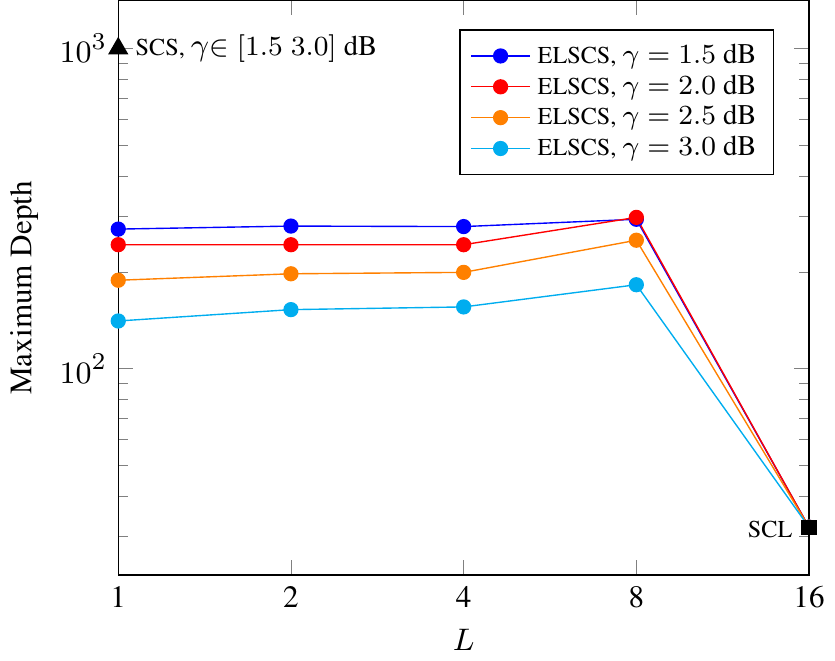}
\caption{Maximum stack depth under different decoding algorithms with different $L$ values and $\gamma$ when $Q=16$, $\delta=12$.}
\end{figure}

\begin{figure}[!htbp]
\centering
\includegraphics{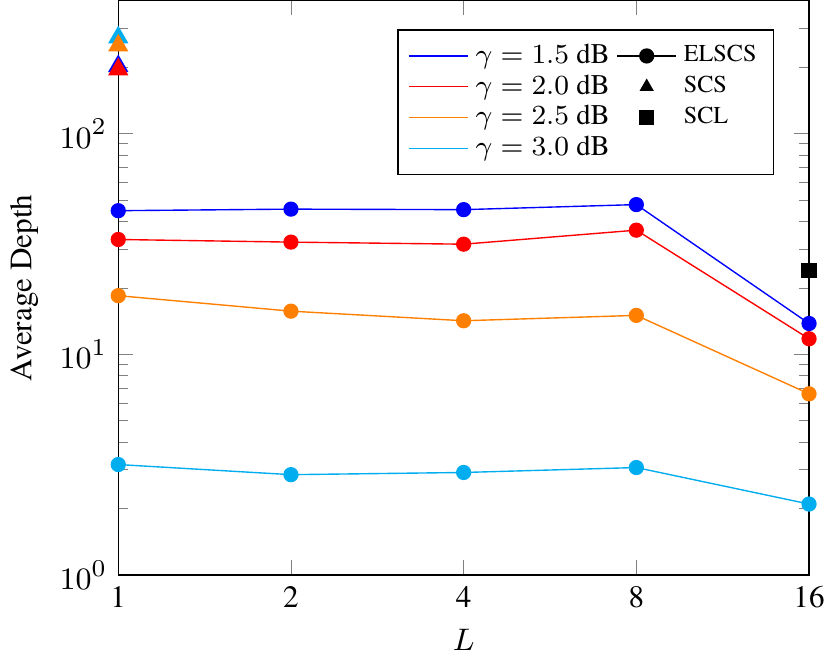}
\caption{Average stack depth under different decoding algorithms with different $L$ values and $\gamma$ when $Q=16$, $\delta=12$.}
\end{figure}

Fig. 12 and Fig. 13 show the statistical results of the maximum stack depth and the average stack depth required by SCS and ELSCS. The stack depth is $1000$ and represents the total depth of stack $\mathcal{A}$ and stack $\mathcal{B}$. We observe that LTPE scheme reduces the required storage size dramatically. Compared with SCS, the maximum stack depth drops $70\%$, from $1000$ to around $300$, which means that a stack with depth of $300$ is enough for the proposed algorithm. At $\gamma$ =$\{1.5$ dB, $2.0$ dB, $2.5$ dB, $3.0$ dB$\}$, the average storage space of ELSCS reduces to $22.27\%$, $15.92\%$, $6.57\%$, $1.19\%$ of that of SCS, respectively. According to analysis in Section \ref{sec:llr}, the value of $|\mathbb{E}\lbrack \mathrm{L}_i\rbrack|$ at unfrozen bit will become larger when $\gamma$ increases. Thus more paths will be pruned because of a large value of $|\mathrm{L}_i|$. This explains why both the maximum and average depths become smaller when $\gamma$ increases. The simulation results corroborate the effectiveness of LTPE scheme in path pruning. We can see that a large amount of storage space can be optimized by using ELSCS decoder for polar codes instead of conventional stack decoder.

\section{Conclusion}
\label{sec:Conclu}

In this paper, we have proposed a complexity-adjustable multimode decoding algorithm ELSCS for polar codes. We firstly study the LLR characteristics of the correct path in decoding process. It was observed that if the decoding bit of the correct path is an unfrozen bit, it happens with a large probability that the path metric difference between two generated paths after path extension is large. A LTPE scheme is designed using this fact to reduce the storage space. Based on the proposed scheme, we employ both the ideas of SCL and SCS to introduce a novel LSCS algorithm, which can combine their complementary advantages. By changing the list size of extending paths, LSCS can adjust the tradeoff between time complexity and computational complexity flexibly while retaining the error performance unchanged. Driven by the low PE utilization problem in LSCS, LSCS is improved to obtain an enhanced version, ELSCS. ELSCS extends two sequential bits at extension stage to make the runtime of different PEs close and reduce the occurrences of some PEs staying idle and waiting. Thus, the time complexity can further decrease.\par
Performance and complexity analyses based on simulations show that without affecting error rate performance, ELSCS can not only reduce storage size but also provide a flexible tradeoff between time and computational complexity. Making use of this property, we can choose different modes of ELSCS algorithm to meet different application requirements at a low computational complexity and computing resource occupancy, thus helping mobile devices in reducing energy consumption as much as possible.

\bibliography{myreference}

\end{document}